\documentstyle[psfig,twoside,fleqn,espcrc2]{article}

\newcommand{\AmS}{{\protect\the\textfont2 
A\kern-.1667em\lower.5ex\hbox{M}\kern-.125emS}}

\title{Common Structures in 2,3 and 4D Simplicial Quantum Gravity
\thanks{Presented by H.S. Egawa}}

\author{H.S.Egawa
        \address{Department of Physics, Tokai University 
        Hiratsuka, Kanagawa 259-12, Japan}
        $^{,}$
        \address{Theory Division, Institute of Particle 
        and Nuclear Studies, KEK, High Energy Accelerator 
        Research Organization, Tsukuba, Ibaraki 305 , Japan}
        ,
        N.Tsuda$^{\,\, {\rm b}}$
        \thanks{supported by Research Fellowship of the Japan 
        Society for the Promotion of Science for Young Scientists.}
        and 
        T.Yukawa 
        \address{Coordination Center for Research and Education, 
        The Graduate University for Advanced Studies, 
        Hayama-cho, Miuragun, Kanagawa 240-01, Japan}
        $^{\! , \,\,{\rm b}}$}
\begin{document}

\begin{abstract}
Two kinds of statistical properties of dynamical-triangulated manifolds 
(DT mfds) have been investigated.
First, the surfaces appearing on the boundaries of 3D DT mfds were 
investigated. 
The string-susceptibility exponent of the boundary surfaces 
($\tilde{\gamma}_{st}$) of 3D DT mfds with $S^{3}$ topology near to the 
critical point was obtained by means of a MINBU (minimum neck baby 
universes) analysis; actually, we obtained $\tilde{\gamma}_{st} \approx -0.5$.
Second, 3 and 4D DT mfds were also investigated by determining the 
string-susceptibility exponent near to the critical point from measuring the 
MINBU distributions.
As a result, we found a similar behavior of the MINBU distributions in 3 and 
4D DT mfds, and obtained $\gamma_{st}^{(3)} \approx \gamma_{st}^{(4)} \approx 0$.
The existence of common structures in simplicial quantum gravity is also 
discussed. 
\end{abstract}

\maketitle 

\section{Introduction} 
The development of simplicial gravity started with the 2D case, and has now 
reached the point of simulating the 4D case.
The scaling structures of boundaries in 2,3 and 4D Euclidian space-time 
using the concept of the geodesic distance have been investigated 
\cite{2_D,3_D,4_D}.
Concerning the analogy of the loop-length distribution in 2D \cite{2_D}, the 
scaling relations of the boundary surface-area distribution in 3D and the 
boundary-volume distribution in 4D have been shown near to the critical point 
\cite{3_D,4_D}. 

First, we are concerned with the surfaces appearing on boundaries in 3D DT 
mfds with $S^{3}$ topology. 
We actually investigated the relation between the boundary surfaces of 3D 
simplicial quantum gravity (SQG) and the random surfaces of 2D SQG.
In order to investigate the statistical quantities of these boundary 
surfaces, we measure the string-susceptibility exponent 
($\tilde{\gamma}_{st}^{(2)}$) of the boundary surface of 3D DT mfds by means 
of a MINBU\cite{Minbu} analysis.
We denote $\tilde{\gamma}_{st}^{(d)}$ as the string-susceptibility exponent 
of the boundary surfaces in $(d+1)$D DT mfds.
Fig.1 shows a schematic picture of the boundary surfaces at a distance $d$ 
in 3 and 4D DT mfds.
\begin{figure}
\centerline{\psfig{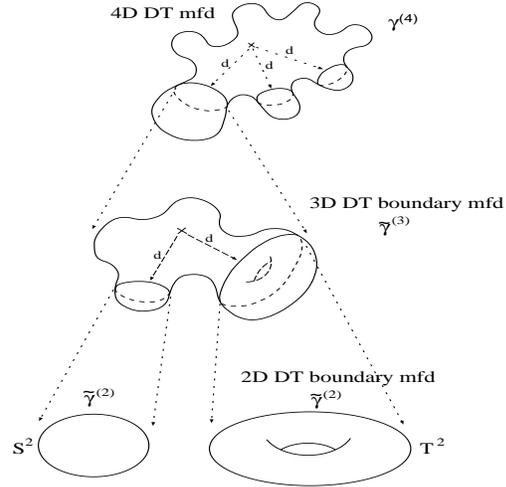}} 
\vspace{-1.0cm}
\caption
{
Schematic picture of an analysis of the boundaries in 3 and 4D 
DT mfds with $\gamma_{st}$ and $\tilde{\gamma}_{st}$. 
}
\label{fig:Boundary}
\vspace{-1.0cm}
\end{figure}
We naively expected equivalence between the boundary surfaces of 3D DT mfds 
and the random surfaces on 2D DT mfds. 

Second, we investigated the structures of 3 and 4D DT mfds \cite{Min_Stru} by 
measuring the standard MINBU distributions, and obtain the 
string-susceptibility exponents ($\gamma_{st}^{(3)}$ and $\gamma_{st}^{(4)}$). 
We found the fact that $\gamma_{st}^{(3)} \approx \gamma_{st}^{(4)} \approx 0$.
The aim of this short paper is to discuss the existence of common 
structures in simplicial quantum gravity. 
%
\section{Models}
%
We start with the Euclidean Einstein-Hilbert action 
in $d$-dimensional SQG, 
\begin{equation}
S_{EH}[\Lambda,G] = \displaystyle{\int} d^{d}\xi \sqrt{g}
(\Lambda-\frac{1}{G}R), 
\end{equation}
where $\Lambda$ is the cosmological constant and $G$ is 
Newton's constant.
We use the lattice action, 
\begin{equation}
S^{(d)}_{Lattice}[\kappa_{d-2},\kappa_{d}] = 
\kappa_{d} N_{d} - \kappa_{d-2} N_{d-2}, 
\end{equation}
where $\kappa_{d-2}$ and $\kappa_{d}$ are coupling constants.
$N_{i}$ denotes the number of $i$-simplexes.
The lattice actions in 2,3 and 4D SQG are as follows: 
\begin{equation}
S^{(2)}[\kappa_{2}] = \kappa_{2} N_{2} \;\;\; (2D), 
\end{equation}
\begin{equation}
S^{(3)}[\kappa_{3},\kappa_{1}] = \kappa_{3} N_{3} 
- \kappa_{1} N_{1} \;\;\; (3D),
\end{equation}
\begin{equation}
S^{(4)}[\kappa_{4},\kappa_{2}] = \kappa_{4} N_{4} 
- \kappa_{2} N_{2} \;\;\; (4D). 
\end{equation}
We consider a partition function, 
\begin{equation}
Z(\kappa_{d-2}, \kappa_{d}) = \sum_{T(d)} 
e^{-S(\kappa_{d-2}, \kappa_{d})}. 
\end{equation}
We sum over all simplicial triangulations ($T(d)$) in order to carry out 
a path integral about the metric. 
%
\section{Analysis of Boundaries}
We now consider the boundary surfaces of 3D DT mfds with $S^{3}$ topology.
The boundary surfaces are located at a distance $d$ from the origin.
The mother boundary surface is defined as the surface with the largest tip 
volume\cite{This_Talk}. 
It is reported that the non-trivial scaling behavior of the surface area 
distribution of the mother boundary surface in 3D SQG has been shown to exist 
near to the critical point\cite{3_D}. 
Therefore, we concentrate on the mother boundary surface for the measurements. 
%
\subsection{String-Susceptibility Exponent of the Mother Boundary in 3D SQG} 
We now show how to measure $\tilde{\gamma}_{st}$ of boundary surfaces having 
different sizes. 
Suppose that the spherical boundary surfaces are obtained; we can then use 
the MINBU algorithm to measure $\tilde{\gamma}_{st}$.
In the case that the topology of the boundary surface is not spherical, but 
is a handle-body, such as a torus, the standard MINBU algorithm does not work 
well. 
Therefore, we have concluded that we should concentrate on spherical boundary 
surfaces.
Fig.2 shows a measurement of the mother spherical boundary 
MINBU distributions of 3D DT mfds with $N_{3}=16K$. 
The lower limit of the volume of the boundary surfaces is $400$.
Our numerical result shows $\tilde{\gamma}_{st}^{(2)} \approx -0.5$.
In 2D SQG with $S^{2}$ topology, the exact calculation of the 
string-susceptibility exponent ($\gamma_{st}^{(2)}$) is well-known to be 
$\gamma_{st}^{(2)} = -0.5$.
We find a correspondence between the string susceptibility of the mother 
spherical boundary surfaces in  3D DT mfds and that of 2D DT mfds.
Therefore, we have obtained evidence of an equivalence between the boundary 
surfaces of 3D SQG and the random surfaces of 2D SQG.
When these boundary surfaces can be recognized as the surfaces of the 
matrix model, 3D DT mfds can be reconstructed by direct products of 
random surfaces of 2D QG and the geodesic distance.
%
\section{String-Susceptibility Exponents in 3 and 4D SQG}
The standard MINBU algorithms are not available for the case of the 
handle-body in an arbitrary dimension. 
It is not easy to choose 3D spherical boundary mfds in 4D DT mfds, because we 
cannot use the Euler's characteristic in 3D in order to distinguish the 
topologies of 3D DT mfds. 
It is therefore difficult to determine $\tilde{\gamma}_{st}^{(3)}$ of the 
boundary mfds in 4D DT mfds by means of the MINBU technique due to the reason 
mentioned above. 
However, we can compare it with the string-susceptibility exponents in 3 and 
4D DT mfds. 
Fig.3 shows the MINBU distributions of 3D DT mfds with $N_{3}=16K$ near to the 
critical point ($\kappa_{0} = 4.09$ and $\kappa_{3} = 2.20$).
We then obtained $\gamma_{st}^{(3)} \approx 0$ of 3D DT mfds by a MINBU 
analysis.
Moreover, the standard Minbu distributions of 4D DT mfds are measured near to 
the critical point; we actually obtained the string-susceptibility exponent 
as $\gamma_{st}^{(4)} \approx 0$. 
%
\section{Summary and Discussions}
The space-time structures in 2 and 3D SQG were investigated using a 
boundary analysis by means of a MINBU analysis.
We have obtained evidence of an equivalence between the boundary surfaces 
of 3D SQG and the random surfaces of 2D SQG with numerical results,  
$\tilde{\gamma}_{st}^{(2)} \simeq \gamma_{st}^{(2)}$.
Furthermore, we obtained the string-susceptibility exponents: both 
$\gamma_{st}^{(3)} \approx 0$ in 3D SQG near to the critical point 
(belonging to the strong-coupling phase) and 
$\gamma_{st}^{(4)} \approx 0$ in 4D SQG near to the critical point 
(belonging to the strong-coupling phase).       
In the 3D case, we conjecture the relation $\tilde{\gamma}_{st}^{(3)} \simeq
\gamma_{st}^{(3)}$, where $\tilde{\gamma}_{st}^{(3)}$ is obtained from the 
boundary mfds in 4D DT mfds. 
If the above conjecture is correct, the higher dimensional SQG 
(i.e., 3 and 4D SQG) can be reduced to the lower dimensional SQG 
(maybe 2D SQG). 
Some similar discussions are found in ref.7. 

\begin{figure}
\centerline{\psfig{file=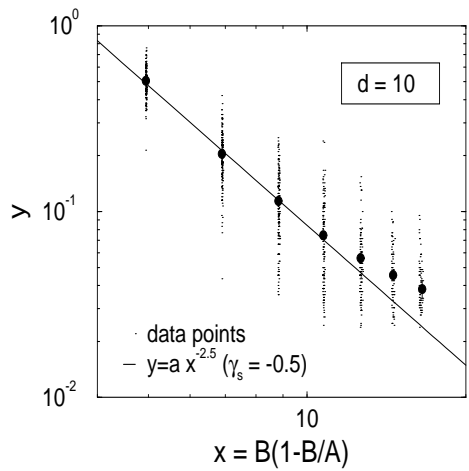,height=6.5cm,width=6.5cm}} 
\vspace{-1.0cm}
\caption
{
Boundary MINBU distributions in terms of the spherical boundary 
surfaces in 3D DT mfds at a distance of 10 with $N_{3}=16K$ 
using $log-log$ scales. The dots represent the raw-data points. 
}
\vspace{-1cm}
\label{fig:Boundary_Minbu_16K_d10}
\end{figure}

\begin{figure}
\centerline{\psfig{file=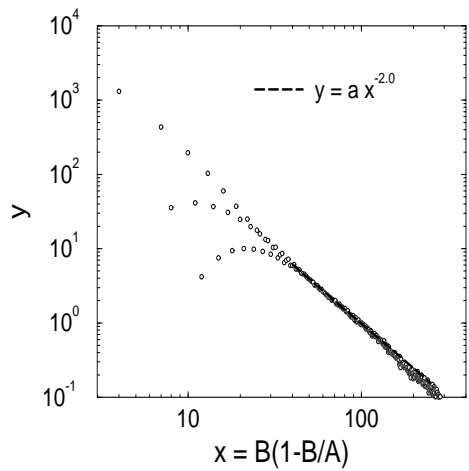,height=6.5cm,width=6.5cm}} 
\vspace{-1cm}
\caption
{
Standard MINBU distributions in 3D DT mfds with $N_{3}=16K$ using $log-log$ 
scales. 
}
\vspace{-0.5cm}
\label{fig:Standard_Minbu_3D_16K_Crit}
\end{figure}

\end{document}